\documentclass[conference]{IEEEtran}
\usepackage{amsmath,amssymb,amsfonts} 
\usepackage{bbm} 
\usepackage{mathrsfs} 
\usepackage{times} 
\usepackage[T1]{fontenc}  

\usepackage{graphicx}
\usepackage{epsfig}
\usepackage{psfrag} 
\usepackage{subfigure} 

\usepackage{textcomp}  
\usepackage{pifont} 

\usepackage{algorithm} 
\usepackage{algorithmic} 

\usepackage{array}
\usepackage{stfloats} 

\usepackage{multicol} 
\usepackage{multirow} 
\usepackage{colortbl} 
\usepackage{booktabs} 
\usepackage{tabularx}  

\usepackage{cite} 
\usepackage{xcolor} 
\usepackage{comment} 

\begin{document}
	\title{Time and Frequency Offset Estimation and Intercarrier Interference Cancellation for \\ AFDM Systems
	}
	\author{Yuankun Tang$^*$,  Anjie Zhang$^*$, Miaowen Wen$^*$, Yu Huang$^\dag$, Fei Ji$^*$, and Jinming Wen$^\ddag$ \\
		{$^*$School of Electronic and Information Engineering, South China University of Technology,  China}\\
		{$^\dag$School of Electronics and Communication Engineering, Guangzhou University, China}\\
		{$^\ddag$Department of Information Science and Technology, Jinan University,  China}\\
		Email: eeyktang@mail.scut.edu.cn, eeanjiezhang@mail.scut.edu.cn, eemwwen@scut.edu.cn,\\ yuhuang@gzhu.edu.cn, eefeiji@scut.edu.cn, jinming.wen@mail.mcgill.ca
		
	}\maketitle              
	\begin{abstract}
        Affine frequency division multiplexing (AFDM) is an emerging multicarrier waveform that offers a potential solution for achieving reliable communications over time-varying channels.	
        This paper proposes two maximum-likelihood (ML) estimators of symbol time offset and carrier frequency offset for AFDM systems. 
        One is called joint ML estimator, which evaluates the arrival time and carrier frequency offset by comparing the correlations of samples. 
        Moreover, we propose the other so-called stepwise ML estimator to reduce the complexity.
        Both proposed estimators exploit the redundant information contained within the chirp-periodic prefix inherent in  AFDM symbols, thus dispensing with any additional pilots. 
        To further mitigate the intercarrier interference resulting from the residual frequency offset, we design a mirror-mapping-based scheme for AFDM systems.
        Numerical results verify the effectiveness of the proposed time and carrier frequency offset estimation criteria and the mirror-mapping-based modulation for AFDM systems.
	\end{abstract}
	\begin{IEEEkeywords}
		Synchronization, affine frequency division multiplexing, time offset estimation, frequency offset estimation, intercarrier interference cancellation.
	\end{IEEEkeywords}
	
	\section{Introduction}
	The sixth-generation (6G) wireless networks are expected to achieve reliable, energy-efficient, and wide coverage communications in high mobility scenarios \cite{6G_IMT2030}.  The high-speed relative motion between transceivers causes the Doppler spread effect, leading to doubly dispersive (time-varying multipath) channels \cite{6GOTFS_Weijie_2023}. Over these channels,  classical multicarrier systems, e.g., the widely employed orthogonal frequency division multiplexing (OFDM), experience substantial performance deterioration caused by intercarrier interference (ICI) \cite{OFDM_Survey}, since Doppler shifts destroy the orthogonality between subcarriers. Against this background, it is urgent to design new waveforms and modulation schemes for achieving reliable communications in high-speed environments \cite{5G_HighSpeedScenario}. 
 
    To cope with the frequency dispersion caused by Doppler shifts in doubly dispersive channels,  orthogonal time-frequency space (OTFS) was proposed to multiplex symbols in the delay-Doppler (DD) domain by the symplectic finite Fourier transform \cite{OTFS_first_2017,OTFS_TimeFrequencySynchronization_Bayat}. The OTFS is a two-dimensional modulation scheme enabling a sparse and stable channel representation in the DD domain. Therefore, it is resilient to time-varying channel effects \cite{OTFS_InterferenceCancellation}. 
    Meanwhile, orthogonal chirp-division multiplexing (OCDM) was proposed as a chirp-based multicarrier waveform \cite{OCDM_first_2016,OCDM_PerformanceAnalysis_2021}, where chirps are complex exponentials with linearly varying instantaneous frequencies. OCDM symbols span the whole bandwidth, such that OCDM systems can collect partial multipath diversity for frequency selective channels, thus outperforming OFDM systems \cite{OCDM_PerformanceAnalysis_2021}. 
    However, the OCDM waveform cannot maximize the diversity in doubly selective channels \cite{Bemani_AFDM_21ICC}.
         
    Affine frequency division multiplexing (AFDM), which is another recently proposed chirp-based waveform,  multiplexes information symbols based on the discrete affine Fourier transform (DAFT) \cite{Bemani_AFDM_21ICC,Bemani_AFDMEqualization_22}.
    The AFDM symbol can be tailored to match the channel profiles through waveform parameter optimization, enabling the separation of all channel paths in the discrete affine Fourier domain. Consequently, AFDM is capable of maximizing both time and frequency diversity gains over doubly selective channels. 
    The bit error rate (BER) performance of AFDM was shown to be better than that of OFDM and the same as that of OTFS for doubly selective channels \cite{Bemani_AFDM_21ICC,Bemani_AFDM_TWC}. Moreover, AFDM has less channel estimation overhead and lower complexity than OTFS \cite{Bemani_AFDM_TWC,AFDMChannelEstimation_ICCC_2022}. Due to its advantages of decoupling delay, the AFDM waveform was studied to sense target information in integrated sensing and communications systems \cite{AFDMwithISAC_ISWC_2022}. Synchronization is a challenge in developing  practical communication systems, especially when dealing with a doubly dispersive channel \cite{TimeFrequencySynchro_2015}. However, to the best of our knowledge, there is no work on the time and frequency offset estimation for the synchronization of AFDM systems so far.

    To fill the gap, this paper proposes time and carrier frequency offset estimation techniques for DAFT-based AFDM systems. Inspired by the maximum-likelihood (ML) estimation scheme that exploits a cyclic prefix (CP) of OFDM systems \cite{TO-CFO_ofdm_MLEsimation}, the proposed methods employ a chirp-periodic prefix (CPP) inherent in  AFDM symbols, dispensing with any pilot overhead.  Unlike the CP, which is a copy of its corresponding symbols, there is a phase difference between the CPP and its corresponding information symbols within AFDM symbols due to the different periodicity of AFDM symbols in the discrete affine Fourier domain.
    Hence, the CPP yields a correlation between the received signals. The received CPP and its corresponding symbols contain information on the time and frequency offset. Based on this observation, we first propose a joint  ML estimator of time and frequency offset by detecting the indices of CPP among samples with a designed length.  
    Then, a stepwise ML estimator is further proposed to reduce the search space with a negligible performance loss.
    Additionally, we propose a mirror-mapping-based modulation scheme for AFDM systems. This modulation scheme exploits the properties of the ICI coefficient, thus mitigating the ICI from  neighboring subcarriers.
    Our simulations study the mean square error (MSE) of the time and frequency offset estimators, which demonstrate that the proposed estimators can satisfy the time and frequency synchronization requirements. BER simulation results confirm the ICI cancellation capability of the mirror-mapping-based AFDM scheme.

        The remainder of this paper is organized as follows. 
        Section~\uppercase\expandafter{\romannumeral2}  describes the   preliminaries of AFDM signals. Section~\uppercase\expandafter{\romannumeral3} proposes two time and carrier frequency offset estimation techniques, followed by the mirror-mapping-based modulation to cancel ICI in Section~\uppercase\expandafter{\romannumeral4}. Numerical results are presented and analyzed in Section~\uppercase\expandafter{\romannumeral5}. Finally, we conclude this paper in Section~\uppercase\expandafter{\romannumeral6}.
	\section{Preliminaries of AFDM}
	In this section, we review preliminary knowledge of AFDM systems. For a detailed description of the AFDM system model and its modulation principles, see \cite{Bemani_AFDM_21ICC}.
	
	Let $\bf{x}$ denote a vector comprising $N$ modulated symbols denoted by $x_m$, $m=0,1,\dots,N-1$, in the discrete affine Fourier domain. These  symbols are mapped into the time domain by inverse DAFT (IDAFT), leading to the modulated signal as 
	\begin{align}\label{IDAFT}
	s\left[ n \right] = \frac{1}{{\sqrt N }}\sum\limits_{m = 0}^{N - 1} x \left[ m \right]{e^{j2\pi \left( {{c_1}{n^2} + {c_2}{m^2} + nm/N} \right)}},
	\end{align}
	where $n=0, 1, \dots, N-1$, $c_1$ and $c_2$ are design parameters. 
    OFDM systems employ the CP, which is the copy of the last several samples of the body of the OFDM symbol,  to construct an equivalent parallel orthogonal channel structure. Similar to this structure, AFDM systems add the CPP instead of the CP due to the different periodicity of the AFDM signal. Based on the periodicity defined in \cite{Bemani_AFDM_21ICC}, the CPP is calculated as
	\begin{align}\label{periodicity} 
	s\left[ n \right] = s\left[ {n + N} \right]{e^{ - j2\pi {c_1}\left( {{N^2} + 2Nn} \right)}},\;\;n =  - L, \dots ,0,
	\end{align}
    where $L$ denotes the length of the CPP. We assume that $L\ge l_{\max}$, with $l_{\max}$ representing the maximum delay of channel paths.
	
    Considering a transmission over the doubly dispersive channel, the received signal is given by
	\begin{align}\label{ReceivedSymbol1}
	r\left[ n \right] = \sum\limits_{l = 0}^\infty  s \left[ {n - l} \right]{g_n}\left( l \right) + w\left[ n \right],
	\end{align}
    where  $w[ n ]\sim\mathcal{CN}( 0,\sigma _n^2)$ denotes the additive white Gaussian noise (AWGN) following a complex Gaussian distribution with a variance of $\sigma _n^2$, ${g_n}\left( l \right)$ is the channel impulse response at instant $n$ and delay $l$, and can be represented by
	\begin{align}\label{ReceivedSymbol2}
	{g_n}\left( l \right) = \sum\limits_{i = 1}^P {{h_i}{e^{ - j2\pi {\alpha_i}n/N}}\delta \left( {l - {l_i}} \right)} ,
	\end{align}
	with ${\delta  (\cdot )}$ denoting the Dirac delta function, $P$ represents the number of channel paths, $\alpha_i$ is the Doppler shift of the $i$-th path normalized by $N$, and $h_i$ and $l_i$ are the complex gain and the path delay associated with the $i$-th path, respectively. Let us set $c_1$ as a positive number not smaller than $\frac{{2{\alpha_{\max }} + 1}}{{2N}}$ with $\alpha_{\max }$ denoting the maximum Doppler shift normalized by $N$, and $c_2$ as any irrational number or a rational number  less than $\frac{{1}}{{2N}}$. Then, when $ 2\alpha_{\max}+l_{\max}+2\alpha_{\max}\l_{\max}<N$, paths with varying delay values or distinct Doppler shifts become distinguishable within the discrete affine Fourier domain. Therefore, the AFDM can obtain full multipath  diversity.
	
\section{Time and Carrier Frequency Offset Estimation}
	In this section, we provide two methods to estimate the symbol time offset and carrier frequency offset of AFDM systems. For the theoretical analysis, we assume that the received signal is only impaired by the AWGN in Sections III and IV. However, we will consider both the doubly dispersive channel and the AWGN channel for numerical simulations in Section V to show the performance of the proposed offset estimation schemes. 
	
	The uncertainties of symbol time and carrier frequency yield the received signal over the AWGN channel as \cite{TO-CFO_ofdm_MLEsimation}
	\begin{align}\label{ReceivedSymbolEsti}
	r\left[ {n } \right] = s\left[ {n  - \theta } \right]{e^{j2\pi \varepsilon  {n } /N}} + w\left[ {n} \right],
	\end{align}    
	where $\theta$ is modeled as a delay representing the integer-valued unknown arrival time of an AFDM symbol, and $\varepsilon$ represents the frequency offset as a fraction of intercarrier spacing normalized with frequency. This carrier frequency offset is caused by the  frequency difference of the local oscillators at the transmitter and  receiver, and hence, all subcarriers have the same $\varepsilon$. 
    Since  $s[ n ]$ results from linear operations on independent identically distributed random variables, it approximates a complex Gaussian process based on the central limit theorem when  $N$ is large enough. However, this process is non-white due to the fact that the presence of a CPP incurs correlations between some sample pairs separated by a distance of $N$ samples, leading to 
     $r[n]$ not being a white process either. Therefore, $r[n]$ contains information on time offset $\theta$ and carrier frequency offset $\varepsilon$, from which we can estimate these uncertain parameters.

	Assume that the receiver samples $2N+L$ consecutive $r[k]$, where $k=1,2,\dots,2N+L$, such that these samples can comprise a complete AFDM symbol with $N+L$ samples long. To estimate the time delay $\theta$ and $\varepsilon$, we need to  find the indices of the CPP of this AFDM symbol denoted by $\mathcal{I}$, where  $\mathcal{I} = \{\theta, \dots, \theta+L-1 \}$. First, we propose the joint  ML estimator, which can be written as
	\begin{align}\label{ML1}
	\left( {\hat \theta ,\hat \varepsilon } \right) &= \mathop {\arg \max }\limits_{\theta ,\varepsilon } \log f\left( {r\left[ 1 \right],r\left[ 2 \right], \dots ,r\left[ {2N + L} \right]} \right) \nonumber\\
	& = \mathop {\arg \max }\limits_{\theta ,\varepsilon } \sum\limits_{k \in {\cal I}} {\log \frac{{f\left( {r\left[ k \right],r\left[ {k + N} \right]} \right)}}{{f\left( {r\left[ k \right]} \right)f\left( {r\left[ {k + N} \right]} \right)}}} ,
	\end{align}    
	where the last equation of (\ref{ML1}) is valid due to the fact that only the samples from $\mathcal{I}$ are pairwise correlated with their corresponding samples  $r(k+N)$, while the other sample pairs are mutually independent.

    In the following, we turn to calculate the two-dimensional distribution ${f\left( {r\left[ k \right],r\left[ {k + N} \right]} \right)}$ for $k \in {\cal I}$. Based on  (\ref{periodicity}), the sample at time $k$ of (\ref{ReceivedSymbolEsti}) can be rewritten as
    \begin{align}\label{AFDM_offset_samples}
    r\left[ k \right] = s\left[ {k + N - \theta } \right]{e^{j2\pi \left( {\varepsilon k/N - \beta_k} \right)}} + w\left[ k \right],
    \end{align} 
    where for brevity we define
     \begin{align}\label{alpha_k }
	{\beta_k} = {c_1}\left( {{N^2} + 2N(k - \theta  - L)} \right).
    \end{align}  
    According to (\ref{ReceivedSymbolEsti}) and (\ref{AFDM_offset_samples}), the expectation of the correlation between  $r\left[ k \right]$ and its corresponding symbol spaced $N$ samples apart $r\left[ k +N\right]$ is given by
    \begin{align}\label{correlation_rk_rk+N}
	\ E\big\{ {r\left[ k \right]{r^*}\left[ {k + N} \right]} \big\} = \left\{ {\begin{array}{*{20}{c}}
{\sigma _s^2{e^{ - j2\pi \left( {\varepsilon  + {\beta _k}} \right)}}}, \; k \in \mathcal{I}\\
0, \; \; \; \; \; \; \; \; \; \; \; \; \; \; \; \; \; \; \; \; \; k \notin \mathcal{I}
\end{array}} \right., 
    \end{align}      
    where $ E\{\cdot\} $ represents the expectation operation, and  $\sigma _s^2 = E\big\{ {{{\left| {s\left[ k \right]} \right|}^2}} \big\}$ denotes the average energy of transmitted symbols.
    Let $\bf{r}$ denote the vector ${\big[ {r\left[ k \right]},{r\left[ {k + N} \right]} \big]^T}$. Hence, we can write $\bf{R}$, i.e., the correlation matrix of $\bf{r}$, as	
    \begin{align}\label{Correlation_matrix_R}
	{\bf{R}} &= \ E\left\{ {{\bf{r}}{{\bf{r}}^H}} \right\} \nonumber\\
 &= \left( {\sigma _s^2 + \sigma _n^2} \right)\left[ 
        {\begin{array}{*{20}{c}}
        1&{\rho {e^{ - j2\pi \left( {\varepsilon  + {\beta _k}} \right)}}}\\
        {\rho {e^{j2\pi \left( {\varepsilon  + {\beta _k}} \right)}}}&1
        \end{array}} \right],
    \end{align}  	
    where $\rho $ represents the magnitude of the correlation coefficient of $r[k]$ and $r[k+N]$. Based on (\ref{correlation_rk_rk+N}), $\rho $ can be calculated as
    \begin{align}\label{AFDM_rho}
	\rho  &=\left| {\frac{ E{\left\{ {r\left[ k \right]{r^*}\left[ {k + N} \right]} \right\}}}{{\sqrt { E\big\{ {{{\left| {r\left[ k \right]} \right|}^2}} \big\} E\big\{ {{{\left| {r\left[ {k + N} \right]} \right|}^2}} \big\}} }}} \right| \nonumber\\
 &= \frac{{\sigma _s^2}}{{\sigma _s^2 + \sigma _n^2}}.
    \end{align} 
    Assume that $\bf{r}$ is a joint Gaussian vector. This assumption can be justified because linear operations on Gaussian random variables result in another Gaussian random variable. 
    It follows~that 
    \begin{align}\label{two_dimensional_distribution1}
	f\left( {r\left[ k \right],r\left[ {k + N} \right]} \right) = \frac{1}{{{\pi ^2}\det ({\bf{R}})}}\exp \left( { - {{\bf{r}}^H}{{\bf{R}}^{ - 1}}{\bf{r}}} \right), \; k \in \mathcal{I}.
    \end{align} 
    Based on (\ref{Correlation_matrix_R}),  (\ref{two_dimensional_distribution1}) can be further calculated as (\ref{two_dimensional_distribution2}) shown at the top of the next page.
    \begin{figure*}[t]
        \normalsize
	\setcounter{equation}{12}
        \begin{align}\label{two_dimensional_distribution2}
		 f\left( {r\left[ k \right],r\left[ {k + N} \right]} \right)  = \frac{1}{{{\pi ^2}{{\left( {\sigma _s^2 + \sigma _n^2} \right)}^2}\left( {1 - {\rho ^2}} \right)}} \exp \left( { - \frac{{{{\left| {r\left[ k \right]} \right|}^2} - 2\rho {\mathop{\rm Re}\nolimits} \left\{ {r\left[ k \right]{r^*}\left[ {k + N} \right]{e^{j2\pi \left( {\varepsilon  + {\beta _k}} \right)}}} \right\} + {{\left| {r\left[ {k + N} \right]} \right|}^2}}}{{\left( {\sigma _s^2 + \sigma _n^2} \right)\left( {1 - {\rho ^2}} \right)}}} \right).
	\end{align}
	\hrulefill
    \end{figure*}
    The one-dimensional complex Gaussian distribution is given by
    \begin{align}\label{one_dimensional_complex_Gaussian_distribution}
    f\left( {r\left[ k \right]} \right) = \frac{1}{{\pi \left( {\sigma _s^2 + \sigma _n^2} \right)}}\exp \left( { - \frac{{{{\left| {r\left[ k \right]} \right|}^2}}}{{\left( {\sigma _s^2 + \sigma _n^2} \right)}}} \right).
    \end{align}      
    Substituting (\ref{two_dimensional_distribution2}) and (\ref{one_dimensional_complex_Gaussian_distribution}) into (\ref{ML1}), the joint ML estimator can be rewritten as
    \begin{align}\label{jointML}
	\left( {\hat \theta ,\hat \varepsilon } \right) = \mathop {\arg \max }\limits_{\theta ,\varepsilon } {\mathop{\rm Re}\nolimits} \left\{ {\gamma \left( \theta  \right){e^{j2\pi \left( {\varepsilon  + {c_1}{N^2}} \right)}}} \right\} - \frac{\rho }{2}\phi \left( \theta  \right),
    \end{align}     
    where $\rm Re\left\{ \cdot \right\}$ represents extracting the real part of a complex number, and we define
    \begin{align}\label{jointMLphigamma}
	\gamma \left( \theta  \right) &= \sum\limits_{k = \theta }^{\theta  + L - 1} {r\left[ k \right]{r^*}\left[ {k + N} \right]{e^{j4\pi {c_1}N\left( {k - \theta  - L} \right)}}}, \\
 \phi \left( \theta  \right) &= \sum\limits_{k = \theta }^{\theta  + L - 1} {{{\left| {r\left[ k \right]} \right|}^2} + {{\left| {r\left[ {k + N} \right]} \right|}^2}}, 
    \end{align}  
    where $\gamma \left( \theta  \right)$ reflects the correlation between sample pairs spaced $N$ samples apart, containing information of the time offset $\theta$ and the frequency offset $\varepsilon$, while  $\phi \left( \theta  \right)$ is an energy term that only depends on the time offset  $\theta$. 

    Furthermore,  to reduce the computational complexity, we propose a stepwise ML estimator that first finds the time offset $\theta$ and then calculates the frequency offset $\varepsilon$ with the estimated $\hat\theta$. To maximize the likelihood function, the optimal $\varepsilon$ is achieved when the term ${\gamma \left( \theta  \right){e^{j2\pi \left( {\varepsilon  + {c_1}{N^2}} \right)}}}$ in (\ref{jointML}) is a real number, and we~obtain
    \begin{align}\label{SepaML}
    \varepsilon \left( \theta  \right) = \frac{{ - \angle \left\{ {\gamma \left( \theta  \right){e^{j2\pi {c_1}{N^2}}}} \right\}}}{{2\pi }} + \epsilon,
    \end{align}
    where $\epsilon$ is an integer and $\angle\{\cdot\}$ denotes the phase of a complex number. Assume that a rough frequency offset has been estimated, leading to $\left| \varepsilon  \right| < 1/2$. Provided that ${\gamma \left( \theta  \right){e^{j2\pi \left( {\varepsilon  + {c_1}{N^2}} \right)}}}$ is a real number, the time offset estimation is given by
    \begin{align}\label{SepaML_time}
	\hat \theta  = \mathop {\arg \max }\limits_\theta  \left| {\gamma \left( \theta  \right)} \right| - \frac{\rho }{2}\phi \left( \theta  \right).
    \end{align}      
     Then, based on the estimated time offset $\hat \theta$, the frequency offset $\varepsilon$ can be estimated via
    \begin{align}\label{SepaML_frequency}
	\hat \varepsilon  = \frac{{ - \angle \left\{ {\gamma \left( \hat \theta  \right){e^{j2\pi {c_1}{N^2}}}} \right\}}}{{2\pi }}.
    \end{align}
\section{ICI Cancellation}
This section first analyzes the property of the ICI coefficient of AFDM systems, based on which a mirror-mapping-based modulation scheme is then proposed to mitigate the ICI caused by the frequency offset. Moreover, we  derive the carrier-to-interference power ratio (CIR) for the plain AFDM and mirror-mapping-based AFDM.

Consider that the received signals are impaired by the residual frequency offset after synchronizing.  Samples in the discrete affine Fourier domain are given by
    \begin{align}\label{ICI_1}
y[\hat m] &= \frac{1}{{\sqrt N }}\sum\limits_{n = 0}^{N - 1} s [n]{e^{j\frac{{2\pi \varepsilon n}}{N}}}{e^{ - j2\pi ({c_2}{{\hat m}^2} + \frac{{\hat mn}}{N} + {c_1}{n^2})}} + W[\hat m] \nonumber\\
{\kern 1pt} & = \sum\limits_{m = 0}^{N - 1} {x[m]{P_{m,\hat m}}{S_{m - \hat m}}}  + W[\hat m], \;\;\hat m = 0, \dots ,N-1,
    \end{align}
where $W[\hat m]$ represents the noise in the discrete affine Fourier domain and
\begin{align}\label{ICI_S}
&{S_q} = \frac{1}{N}\sum\limits_{n = 0}^{N - 1} {{e^{j\frac{{2\pi n}}{N}(q + \varepsilon )}}} = \frac{{\sin \left( {\pi \varepsilon } \right){e^{j\pi \left( {\varepsilon \left( {1 - \frac{1}{N}} \right) - \frac{{q}}{N}} \right)}}}}{{N\sin \left( {\frac{\pi }{N}\left( {q + \varepsilon } \right)} \right)}}, \\
&P_{m,\hat m}= {e^{j2\pi c_2({m^2} - {{\hat m}^2})}}.
\end{align}
We define ${Q_{m,\hat m}} = {P_{m,\hat m}}{S_{m - \hat m}}$ as the ICI coefficient of AFDM systems similar to that of OFDM systems \cite{Mirror-mapping-modulation_ICI-concellation}.

  \begin{figure}[t]
	   \centering  

			\subfigure[]{
				
				\includegraphics[width=3.5in]{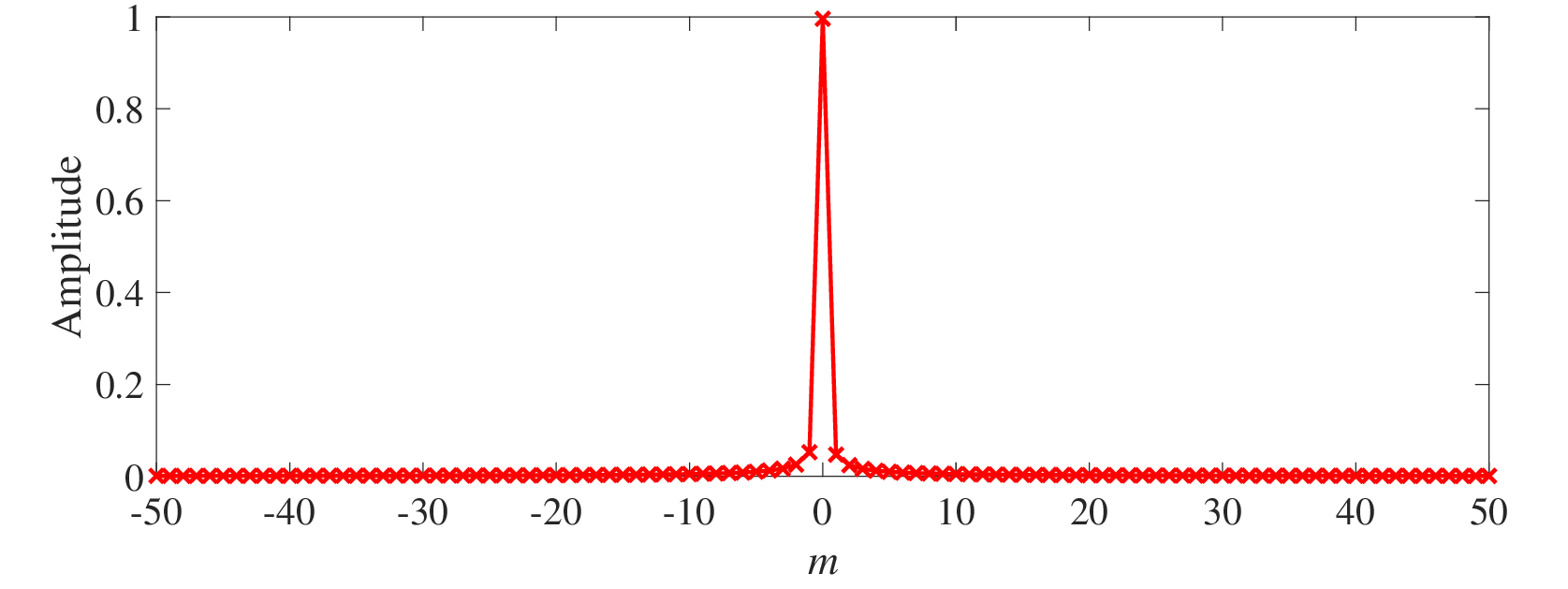}  
			}
			\subfigure[]{
				 
				\includegraphics[width=3.5in]{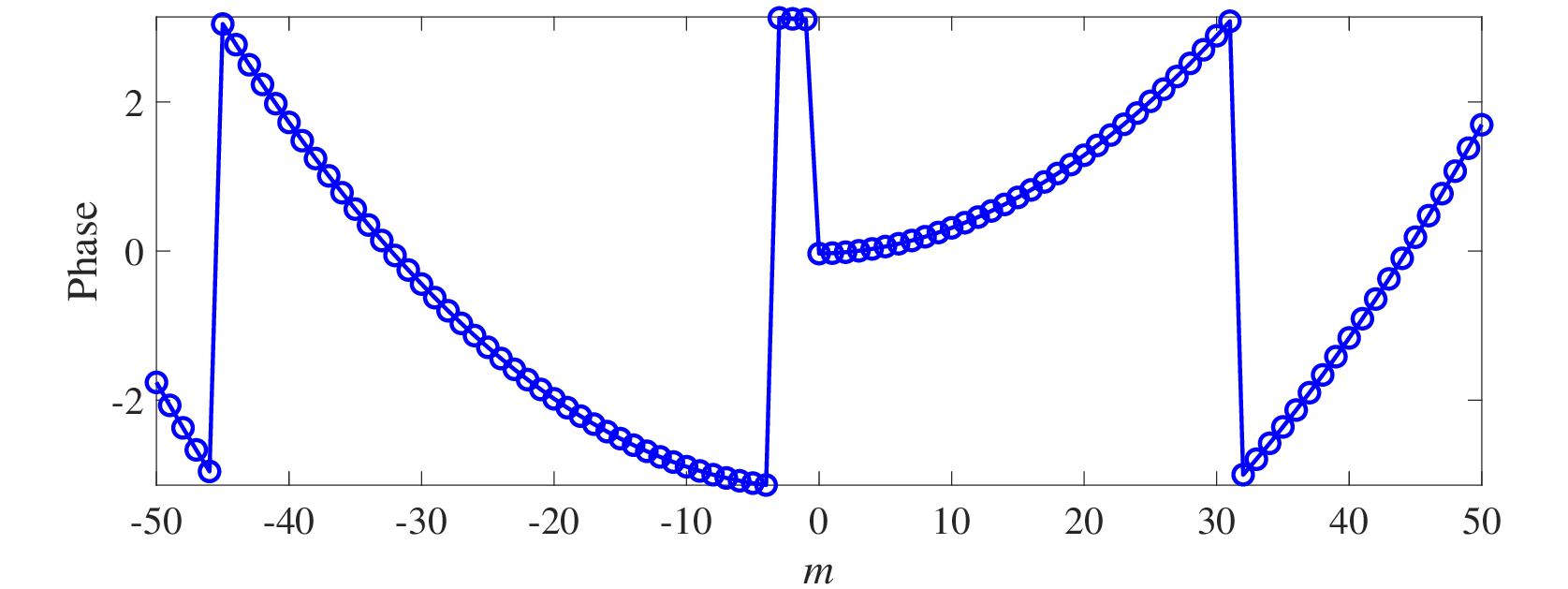}
			}
	
		\caption{Amplitude (a) and phase (b) of the ICI coefficient ${Q_{m,\hat m}}$.}    
		\label{ICI_coefficient}    
	\end{figure}
Figure \ref{ICI_coefficient} presents the amplitude and phase of the ICI coefficient ${Q_{m,\hat m}}$ when $\hat m =0$, $N=1024$, $c_2 = \frac{1}{2N}$, and $\varepsilon = 0.05$. The ICI coefficient ${Q_{m,\hat m}}$ is a periodic function with period $N$. It can be observed that the  amplitude of ${Q_{m,0}}$ is the largest at $m=0$ and dramatically decreases to zero for large values of $\left| m \right|$, which means that ICI mainly comes from the neighboring subcarriers. Furthermore, for small values of $\left| m \right|$, $\left| {\angle {Q_{m,0}}} \right| \approx \pi $ when $m<0$ and  $ {\angle {Q_{m,0}}}  \approx 0$ when $m>0$. This property implies that ${ {Q_{m,0}}}\approx -{ {Q_{-m,0}}}$, based on which we design the mirror-mapping-based scheme in the sequel.

	\begin{figure}[t]
		\centering
		{\includegraphics[width=3.45in]{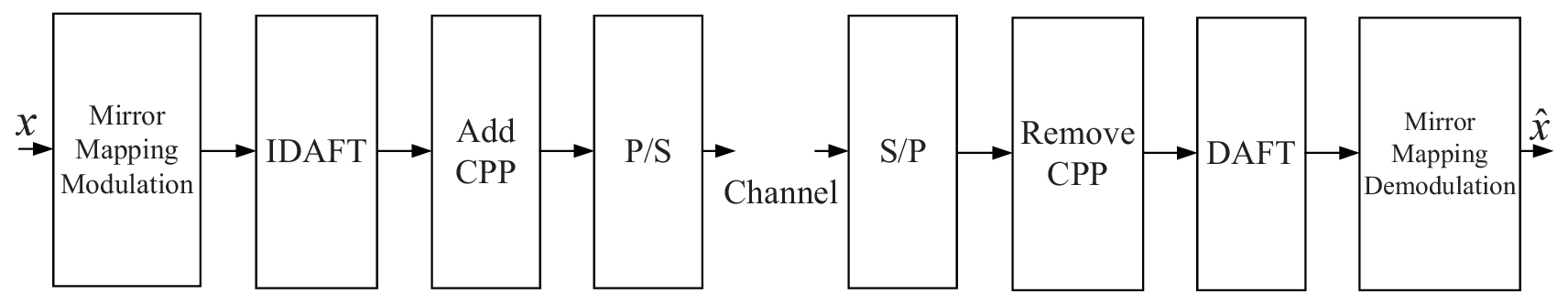}}
		\caption{Schematic diagram of mirror-mapping-based AFDM systems.}
		\label{fig1}
	\end{figure}
The block diagram of the mirror-mapping-based AFDM system is illustrated in Fig. \ref{fig1}. For the mirror-mapping-based modulation, the input modulated symbols are first grouped into blocks. Each block contains $(N/2-1)$ symbols. Then, these  symbols are mapped into the actual transmitted symbols $\tilde x[m]$ on $N$ subcarriers via the one-to-two mirror-mapping-based criterion:
\begin{align}\label{ICI_3}
\tilde x[m] = \left\{ \begin{array}{l}
0, {\kern 1pt} {\kern 1pt}  {\kern 1pt} {\kern 1pt} {\kern 1pt} {\kern 1pt} {\kern 1pt} {\kern 1pt} {\kern 1pt} {\kern 1pt} {\kern 1pt} {\kern 1pt} {\kern 1pt} {\kern 1pt} {\kern 1pt} {\kern 1pt} {\kern 1pt} {\kern 1pt} {\kern 1pt} {\kern 1pt} {\kern 1pt} {\kern 1pt} {\kern 1pt} {\kern 1pt} {\kern 1pt} {\kern 1pt} {\kern 1pt} {\kern 1pt} {\kern 1pt} {\kern 1pt} {\kern 1pt} {\kern 1pt} {\kern 1pt} {\kern 1pt} {\kern 1pt} {\kern 1pt} {\kern 1pt} {\kern 1pt} {\kern 1pt} {\kern 1pt} {\kern 1pt} {\kern 1pt} {\kern 1pt} {\kern 1pt} {\kern 1pt} m = 0,N/2\\
x[m],{\kern 1pt} {\kern 1pt} {\kern 1pt} {\kern 1pt} {\kern 1pt} {\kern 1pt} {\kern 1pt} {\kern 1pt} {\kern 1pt} {\kern 1pt} {\kern 1pt} {\kern 1pt} {\kern 1pt} {\kern 1pt} {\kern 1pt} {\kern 1pt} {\kern 1pt} {\kern 1pt} {\kern 1pt} {\kern 1pt} {\kern 1pt} {\kern 1pt} {\kern 1pt} {\kern 1pt} {\kern 1pt} {\kern 1pt} {\kern 1pt} {\kern 1pt} {\kern 1pt} {\kern 1pt} m = 1,2,...,N/2 - 1{\kern 1pt} \\
 - x[N - m],m = N/2 + 1,...,N - 1
\end{array} \right.,
\end{align}
where the $0$-th and $N/2$-th subcarriers are null to satisfy the opposite polarity condition. Note that the signal on the $m$-th subcarrier, $m=1, 2,\dots,N/2-1$, conveys the same  data information as its corresponding signal on the $(N-m)$-th subcarrier.
The mirror-mapping-based demodulation combines the received signals on the pair of the $m$-th and $(N-m)$-th subcarriers via
    \begin{align}\label{EGC}
\hat x[\hat m] = \frac{1}{2}\left( {y[\hat m] - y[N - \hat m]} \right).
    \end{align}
 
 	\begin{figure}[t]
 	\centering
 	{\includegraphics[width=2.6in]{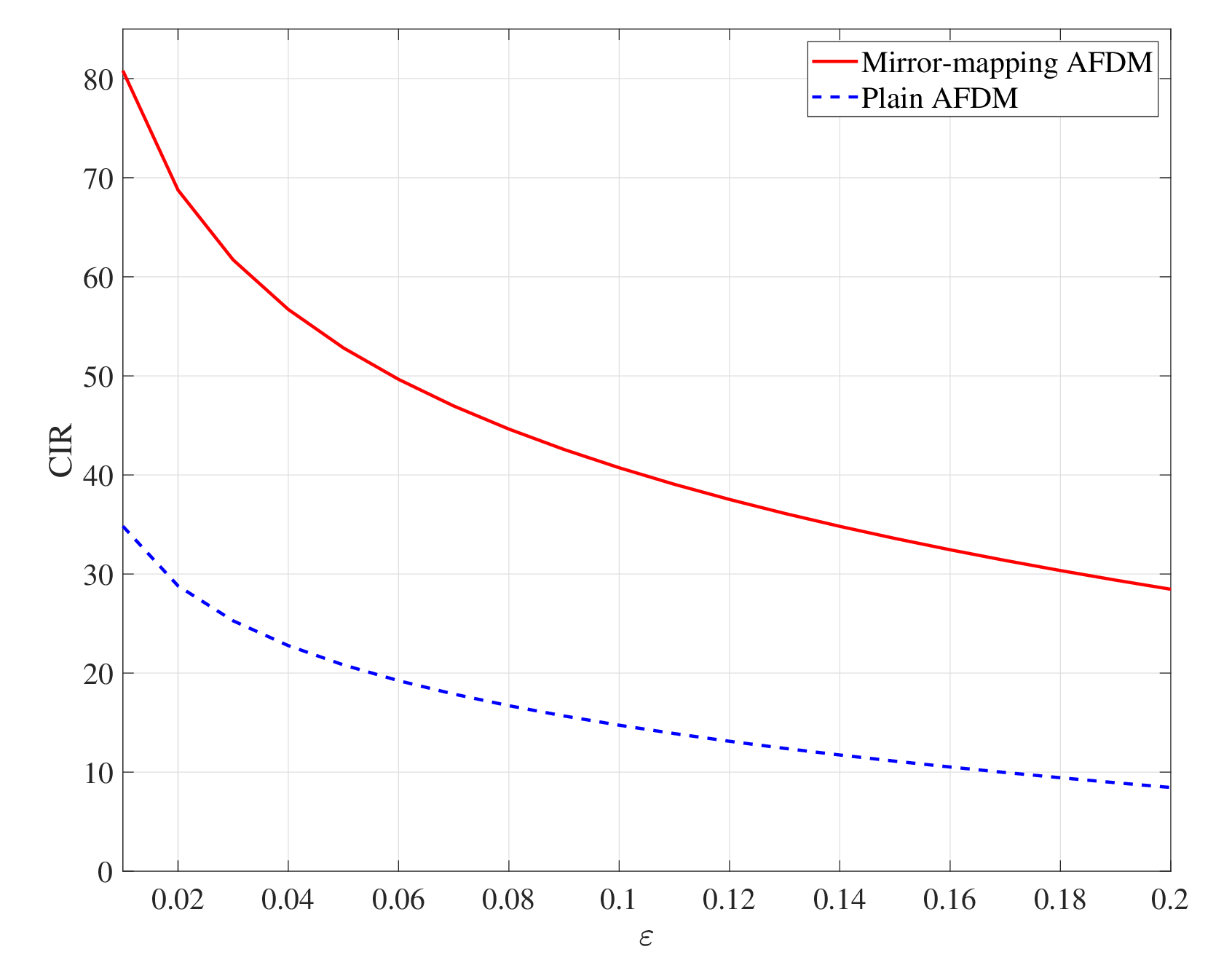}}
 	\caption{CIR comparison of the plain AFDM and the mirror-mapping-based AFDM versus frequency offset $\varepsilon$.}
 	\label{fig_CIR}
 \end{figure}   
Then, we use the CIR to evaluate the  ICI level without considering the noise power.
Based on (\ref{ICI_1}), the CIR of the plain AFDM is given by
    \begin{align}\label{CIR1}
{\rm{CIR}_{{\rm{AFDM}}}} = \frac{{|{S_0}{{\rm{|}}^2}}}{{\sum\limits_{m = 1,m \ne \hat m}^{N - 1} {{{\left| {{P_{m,\hat m}}{S_{m - \hat m}}} \right|}^2}} }}.
    \end{align}
The combined signals via the mirror-mapping-based demodulation  can be rewritten as
    \begin{align}\label{EGC2}
\hat x[\hat m]  &= (2{S_0}{{ - }}{S_{ - 2{{\hat m}}}}{\rm{ - }}{S_{{{2\hat m}}}})x[\hat m] \nonumber\\
&+ \hspace{-0.1cm}\sum\limits_{m = 1,m \ne \hat m}^{N/2 - 1}\hspace{-0.2cm} {{P_{m,\hat m}}}\hspace{-0.1cm} \left( {{S_{m - \hat m}} \hspace{-0.1cm}+ \hspace{-0.1cm}{S_{\hat m - m}}\hspace{-0.1cm} -\hspace{-0.1cm} {S_{m + \hat m}} \hspace{-0.1cm}-\hspace{-0.1cm} {S_{ - m - \hat m}}} \right)x[m]\nonumber\\& + W[\hat m] - W[N - \hat m].
    \end{align}
Thus, the CIR of the mirror-mapping-based scheme can be calculated as
    \begin{align}\label{CIR2}
&{\rm{CI}}{{\rm{R}}_{{\rm{MM}}}}\hspace{-0.05cm} = \frac{2}{{N - 2}}\sum\limits_{\hat m = 1}^{N/2 - 1} 
\nonumber\\
&\hspace{+0.2cm}{\frac{{|2{S_0}{\rm{ - }}{S_{ - 2{{\hat m}}}}{{ - }}{S_{{{2\hat m}}}}{{\rm{|}}^2}}}{\sum\limits_{m = 1,m \ne \hat m}^{N/2 - 1}\hspace{-0.2cm}  \left|{{P_{m,\hat m}}}\hspace{-0.1cm}\left( {{S_{m - \hat m}} \hspace{-0.1cm}+ \hspace{-0.1cm}{S_{\hat m - m}}\hspace{-0.1cm} -\hspace{-0.1cm} {S_{m + \hat m}} \hspace{-0.1cm}-\hspace{-0.1cm} {S_{ - m - \hat m}}} \right)\right|^2}}.
    \end{align}

Figure \ref{fig_CIR} compares the CIR of the plain AFDM and that of mirror-mapping-based AFDM for different values of $\varepsilon$, considering $N=1024$ and $c_2 = \frac{1}{2N}$. We can observe that the mirror-mapping-based AFDM presents much better performance than the plain AFDM, which can be explained by the fact that the interference from neighboring subcarriers is sufficiently suppressed by the mirror-mapping-based scheme.

    \begin{figure}[t]
    \centering
    {\includegraphics[width=2.6in]{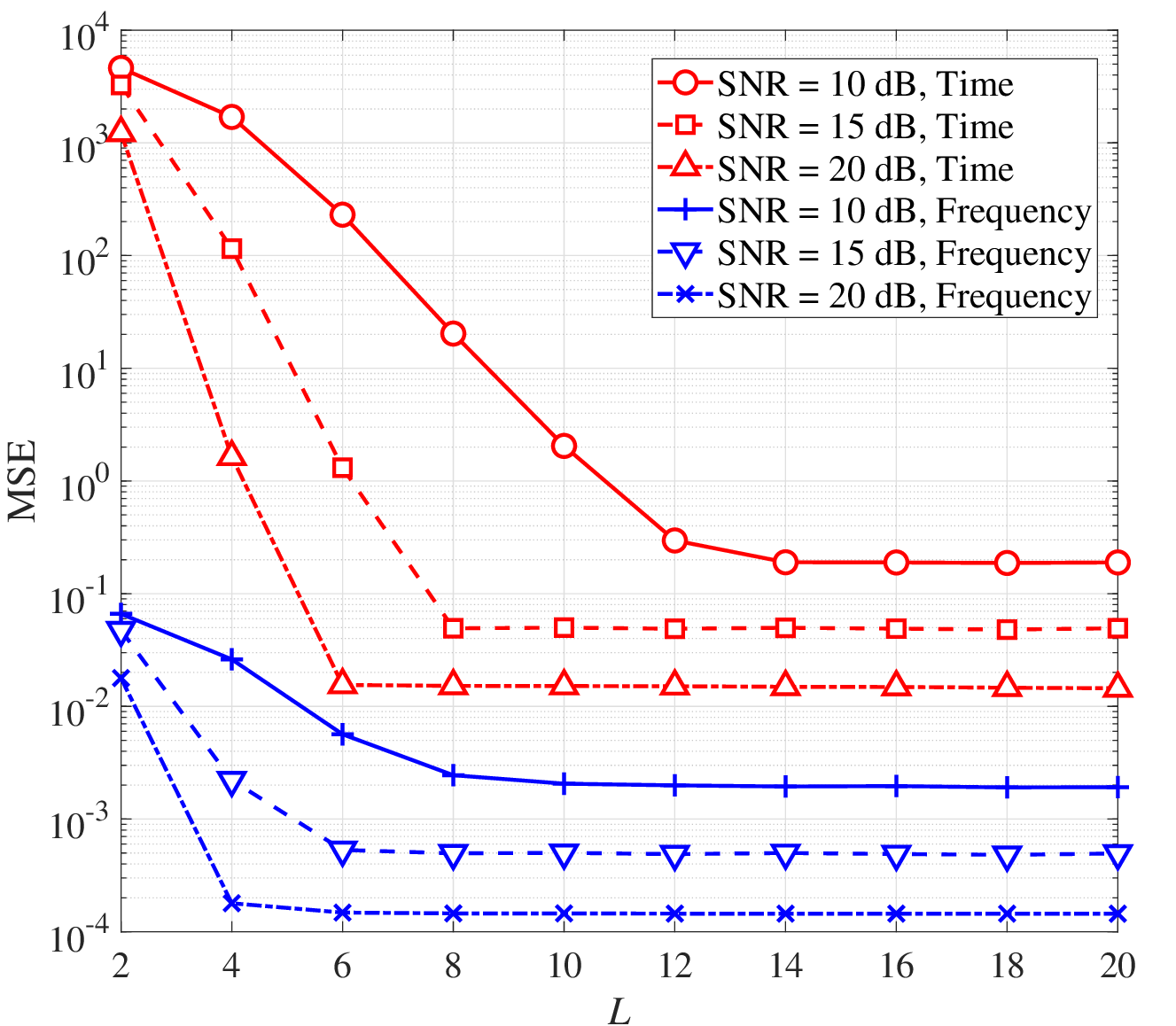}}
    \caption {Performance of time and frequency offset estimators versus the length $L$ of CPP  for AWGN channels employing the joint ML criterion with $\text{SNR} = 10,15,\; \text{and}\; 20 \text{dB}$.}
    \label{AFDM_fig2}
    \end{figure}
\section{Numerical Simulations}
    In this section, we first study the effect of the CPP length $L$ and SNR on the performance of the proposed two ML estimators for AWGN and doubly dispersive channels. To evaluate the estimator performance, we employ the MSE of the estimated time and frequency offset. 
    The simulated time offset and carrier frequency offset are shown using the normalized sample interval and normalized intercarrier space, respectively. Then, the proposed mirror-mapping-based AFDM is compared with the plain AFDM in terms of BER to show its ICI cancellation capability.
    The binary phase shift keying (BPSK) modulation is employed.
    We choose the following parameters: $N=256$ and $c_2=\frac{1}{2N}$. For doubly dispersive channels, we consider the maximum delay of channel path $l_{\max}=1$, the maximum normalized Doppler shift $\alpha_{\max}=2$, and  the number of channel paths $P=5$. 
    
\begin{figure}[t]
    \centering
    {\includegraphics[width=2.7in]{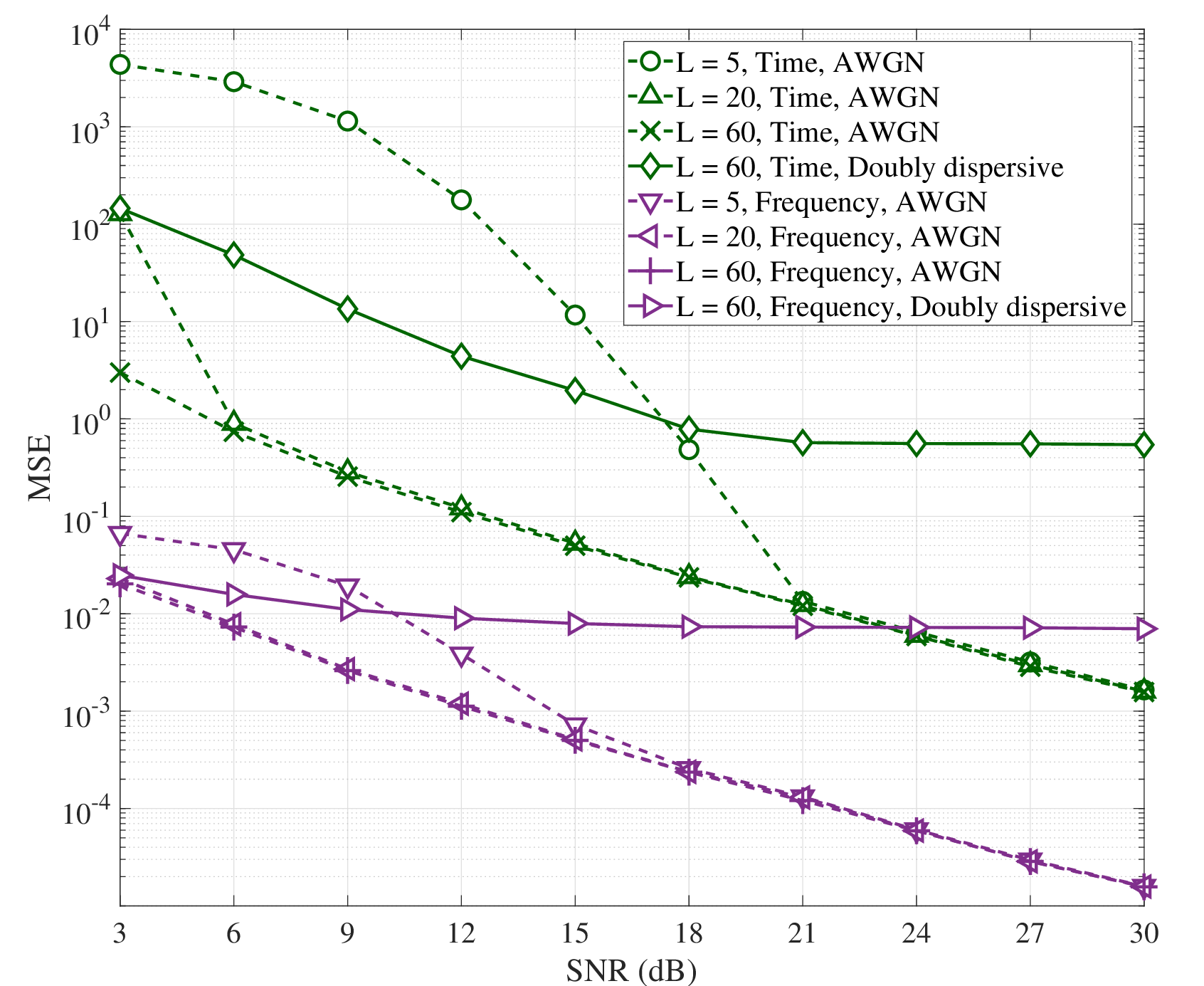}}
    \caption {MSE performance of time and frequency offset estimators versus SNR for AWGN and doubly dispersive channels employing the joint ML criterion with the length of CPP $L = 5,20,\; \text{and}\; 60$.}
    \label{AFDM_fig3}
    \end{figure}

    Figure \ref{AFDM_fig2} presents the joint ML estimator performance versus $L$ for AWGN channels with $\text{SNR} = 10,15,\; \text{and}\; 20 \text{dB}$. We can observe that there are error floors of  both time and carrier frequency offset, which can be achieved as long as the value of $L$ exceeds a specific threshold.     
    Moreover, this threshold of $L$ will decrease with the rise of SNR.  
    Furthermore, it can be observed that the dotted curves in Fig. \ref{AFDM_fig3} will overlap perfectly with the increase of SNR, which is consistent with the threshold phenomenon of $L$ in  Fig. \ref{AFDM_fig2}.
    These results infer that if the value of  $L$ surpasses the threshold determined by SNR, increasing $L$ cannot bring any performance gain of estimating the time and frequency offset for AWGN channels.
    
    In Fig. \ref{AFDM_fig3}, we show the MSE of the estimated time and frequency offset versus SNR, where AWGN and doubly dispersive channels are considered. We use the joint ML criterion with the length of CPP $L = 5,20,\; \text{and}\; 60$. It can be observed that the MSE performance of time and frequency offset under dispersive channels is worse than that with the same CPP length under AWGN channels. Moreover, an error floor appears for doubly dispersive channels  but not for AWGN channels. These results can be explained by the fact that doubly dispersive channels have a complex correlation structure, where the path delay and Doppler shifts  cause  the signals of CPP and their corresponding symbol separated by $N$ samples to be impaired by different interference. However,  the time estimation error for doubly dispersive channels is less than one percent of AFDM symbol interval, provided $\text{SNR} \ge 15 \text{dB}$. Hence, the time estimator is eligible to generate a stable clock \cite{Book.DigitalCommunicatiions.Proakis.2008}.

    \begin{figure}[t]
    \centering
    {\includegraphics[width=2.7in]{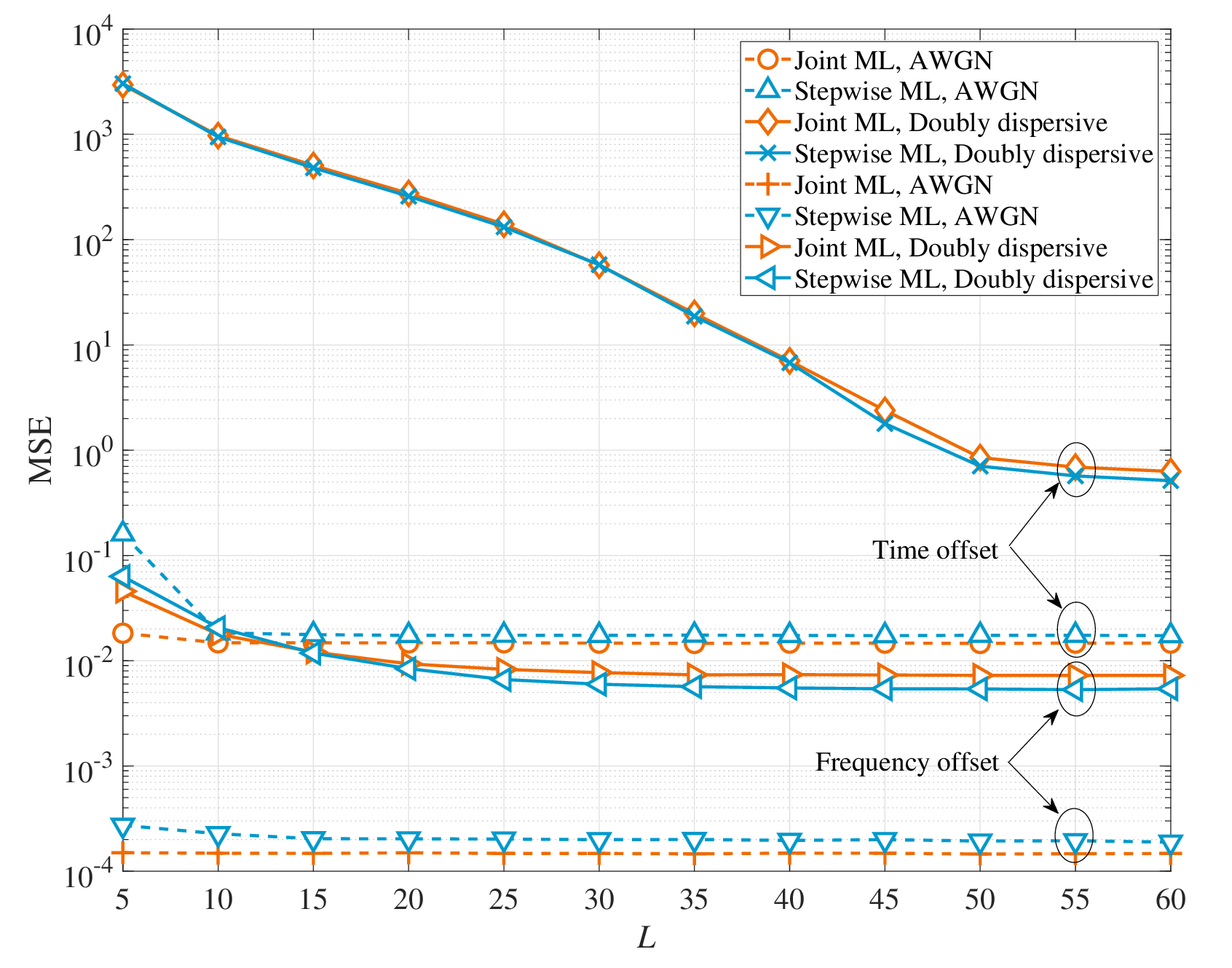}}
    \caption {MSE performance of time and frequency offset estimators versus the length $L$ of CPP for AWGN and doubly dispersive channels with $\text{SNR} = 20 \text{dB}$ employing joint ML and stepwise ML criteria.}
    \label{AFDM_fig4}
    \end{figure}    
    Figure \ref{AFDM_fig4} compares the MSE performance of the joint ML estimator and the stepwise ML estimator over AWGN and doubly dispersive channels  for $\text{SNR} = 20 \text{dB}$ and different lengths of CPP $L$. As shown in Fig. \ref{AFDM_fig4},  the stepwise ML estimator shows similar performance but with lower complexity to the joint ML estimator. However, it outperforms the joint ML estimator for doubly dispersive channels. This counterintuitive result can be accounted for by the fact that  the ML estimators operate under doubly dispersive channels for which these estimators are not customized.
    
    \begin{figure}[t]
    \centering
    {\includegraphics[width=2.7in]{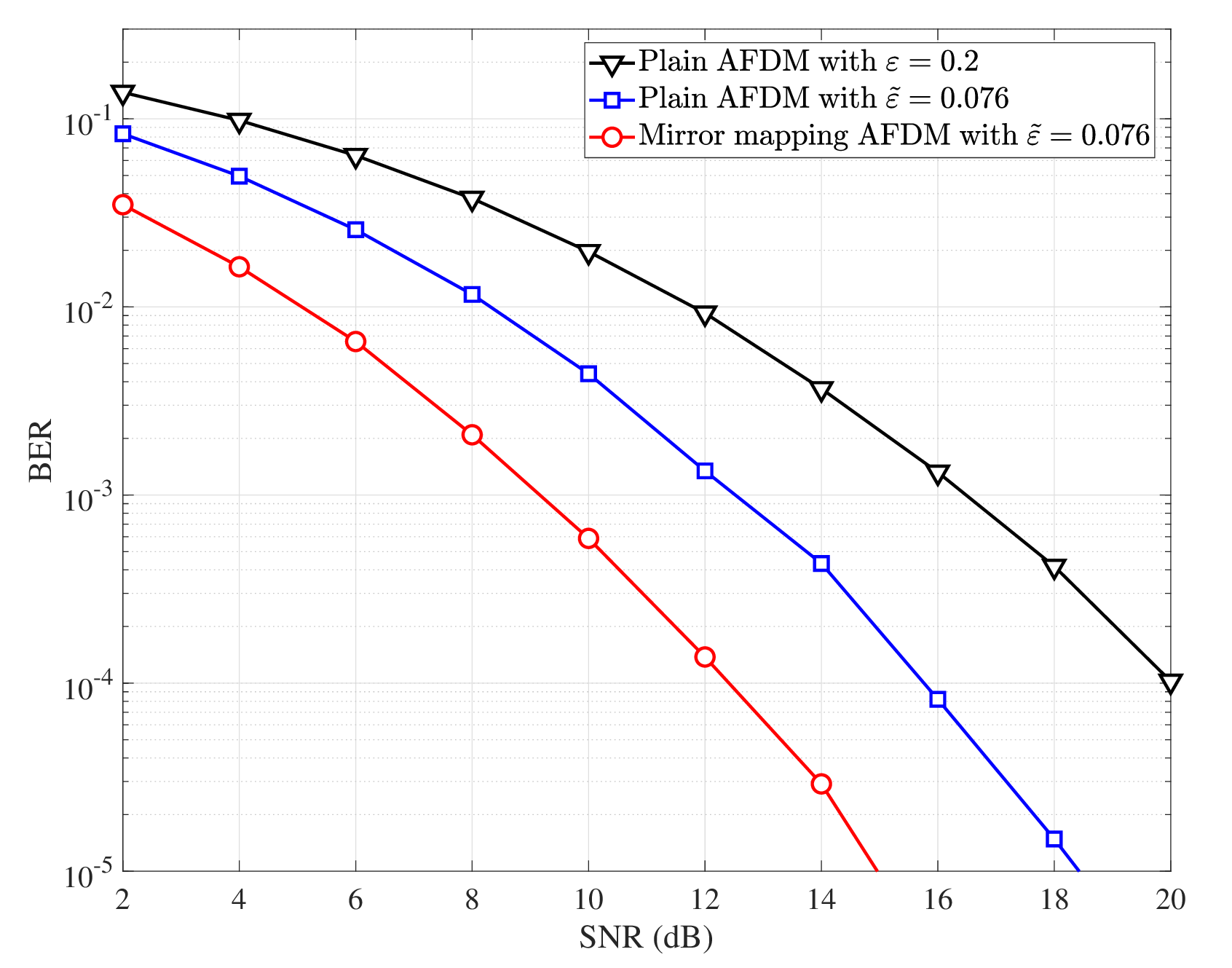}}
    \caption {BER comparison between the plain AFDM, mirror-mapping-based AFDM, and mirror-mapping-based AFDM with the stepwise ML estimator for doubly dispersive channels and $L=30$.}
    \label{AFDM_ICI_BER}
    \end{figure}
	 To further mitigate the ICI caused by the residual carrier frequency offset after synchronization, mirror-mapping-based AFDM is employed. Its  BER performance and that of the plain AFDM are presented in Fig. \ref{AFDM_ICI_BER}. We consider the frequency offset $\varepsilon = 0.2$, $L=30$, and doubly dispersive channels. The residual carrier frequency offset after employing the stepwise ML estimator, denoted by $\tilde{\varepsilon}$,  is assumed to be $\tilde{\varepsilon}=0.076$. To maintain the same spectral efficiency for all schemes, plain AFDM uses $N/2$ BPSK data symbols to occupy half of the total subcarriers, and each data symbol is surrounded by two null subcarriers \cite{Mirror-mapping-modulation_ICI-concellation}. As expected, the mirror-mapping-based AFDM shows lower BER than plain AFDM, which confirms the ICI cancellation ability of the mirror-mapping-based modulation for AFDM systems. 
    
\section{Conclusion}
In this paper, we have proposed two time and carrier frequency offset estimators for the DAFT-based AFDM systems. We have derived the joint ML estimation criterion and developed the stepwise ML estimator to reduce the complexity. In addition, we have proposed the mirror-mapping-based modulation scheme for AFDM systems. Simulation results have demonstrated that these ML criteria can accurately  estimate the time and frequency offset for doubly dispersive channels, and that the mirror-mapping-based AFDM can suppress the ICI resulting from the residual frequency offset.
\section*{Acknowledgment}
This work was supported in part by National Key R\&D Program of China under Grant 2023YFB2904500, and in part by the Guangdong Basic and Applied Basic Research Foundation under Grant 2021B1515120067.
\bibliographystyle{IEEEtran} 
\bibliography{IEEEabrv,ref_AFDM}
\end{document}